\documentclass[twocolumn,superscriptaddress,floatfix,prl]{revtex4}
\usepackage[latin9]{inputenc}
\usepackage{esint}
\usepackage{graphicx}
\usepackage{units}

\makeatletter
\@ifundefined{textcolor}{}
{%
 \definecolor{BLACK}{gray}{0}
 \definecolor{WHITE}{gray}{1}
 \definecolor{RED}{rgb}{1,0,0}
 \definecolor{GREEN}{rgb}{0,1,0}
 \definecolor{BLUE}{rgb}{0,0,1}
 \definecolor{CYAN}{cmyk}{1,0,0,0}
 \definecolor{MAGENTA}{cmyk}{0,1,0,0}
 \definecolor{YELLOW}{cmyk}{0,0,1,0}
}

\makeatother

\begin{document}

\title{Charge induced nematicity in FeSe}

\author{Pierre Massat }

\affiliation{Laboratoire Matériaux et Phénomènes Quantiques, UMR 7162 CNRS, Université
Paris Diderot, Bat. Condorcet 75205 Paris Cedex 13, France}

\author{Donato Farina}

\affiliation{Laboratoire Matériaux et Phénomènes Quantiques, UMR 7162 CNRS, Université
Paris Diderot, Bat. Condorcet 75205 Paris Cedex 13, France}

\author{Indranil Paul}

\affiliation{Laboratoire Matériaux et Phénomènes Quantiques, UMR 7162 CNRS, Université
Paris Diderot, Bat. Condorcet 75205 Paris Cedex 13, France}

\author{Sandra Karlsson}

\affiliation{Institut N\'e\'el, CNRS UPR2940 and Universit\'e Grenoble Alpes, 38042 Grenoble, France}

\author{Pierre Strobel}

\affiliation{Institut N\'e\'el, CNRS UPR2940 and Universit\'e Grenoble Alpes, 38042 Grenoble, France}

\author{Pierre Toulemonde}

\affiliation{Institut N\'e\'el, CNRS UPR2940 and Universit\'e Grenoble Alpes, 38042 Grenoble, France}

\author{Marie-Aude M\'easson}

\affiliation{Laboratoire Matériaux et Phénomènes Quantiques, UMR 7162 CNRS, Université
Paris Diderot, Bat. Condorcet 75205 Paris Cedex 13, France}

\author{Maximilien Cazayous}

\affiliation{Laboratoire Matériaux et Phénomènes Quantiques, UMR 7162 CNRS, Université
Paris Diderot, Bat. Condorcet 75205 Paris Cedex 13, France}

\author{Alain Sacuto}

\affiliation{Laboratoire Matériaux et Phénomènes Quantiques, UMR 7162 CNRS, Université
Paris Diderot, Bat. Condorcet 75205 Paris Cedex 13, France}

\author{Shigeru Kasahara}

\affiliation{Department of Physics, Kyoto University, Kyoto 606-8502, Japan}

\author{Takasada Shibauchi}

\affiliation{Department of Advanced Materials Science, University of Tokyo, Kashiwa, Chiba 277-8561, Japan}

\author{Yuji Matsuda}

\affiliation{Department of Physics, Kyoto University, Kyoto 606-8502, Japan}

\author{Yann Gallais}
\email{yann.gallais@univ-paris-diderot.fr}
\affiliation{Laboratoire Matériaux et Phénomènes Quantiques, UMR 7162 CNRS, Université
Paris Diderot, Bat. Condorcet 75205 Paris Cedex 13, France}

\date{\today }

\begin{abstract}
The spontaneous appearance of nematicity, a state of matter that breaks rotation but not translation symmetry, is one of the most intriguing property of the iron based superconductors (Fe SC), and has relevance for the cuprates as well. Establishing the critical electronic modes behind nematicity remains however a challenge, because their associated susceptibilities are not easily accessible by conventional probes. Here using FeSe as a model system, and symmetry resolved electronic Raman scattering as a probe, we unravel the presence of critical charge nematic fluctuations near the structural / nematic transition temperature, T$_S\sim$ 90 K. The diverging behavior of the associated nematic susceptibility foretells the presence of a Pomeranchuk instability of the Fermi surface with d-wave symmetry. The excellent scaling between the observed nematic susceptibility and elastic modulus data demonstrates that the structural distortion is driven by this d-wave Pomeranchuk transition. Our results make a strong case for charge induced nematicity in FeSe.
\end{abstract}

\maketitle

Electronic nematicity, whereby electrons break rotational symmetry spontaneously, is a ubiquitous property of the iron-based superconductors (Fe SC) \cite{Fisher10}. As it is often accompanied by magnetic order, an established route to nematicity is via critical magnetic fluctuations \cite{Fernandes14}. However, this mechanism has been questionned in the iron-chalcogenide FeSe, where the nematic transition occurs without magnetic order, indicating a different paradigm for
nematicity \cite{McQueen09,Imai09,Baek14, Bohmer15}.
\par
Despite its simple crystallographic structure, FeSe displays remarkable properties. Its superconducting transition
temperature $T_c$ is relatively low at ambient pressure ($\sim$9~K), but it reaches up to 37~K upon applying hydrostatic
pressure \cite{Medvedev09,Garbarino09}. Its Fermi energy is small~\cite{Kasahara14, Terashima14,Watson15,Audouard15}, and in the normal state it shows bad metal
behavior~\cite{Aichhorn10,Kasahara14}. Its nematic properties are peculiar as well.
The lattice distortion, elastic softening and elasto-resistvity measurements associated with the structural transition at $T_S \sim$ 90~K are comparable with other Fe SC \cite{McQueen09,Bohmer15,Watson15}, yet Nuclear Magnetic Resonance (NMR) and inelastic neutron scattering measurements do not detect sizable low energy spin
fluctuations above $T_S$ \cite{Imai09,Bohmer15,Wang15}, putting into question the spin nematic scenario envisaged in other Fe SC \cite{Fernandes14}. While it has been argued that the magnetic scenario may still apply \cite{Glasbrenner15,Wang-Kivelson15,Chubukov15,Yu15,Kontani15}, there is growing interest in alternative scenario where charge or orbital degrees of freedom play a more pre-dominant role than spins \cite{Baek14,Su15,Watson15,Jiang15}. However, until now direct experimental observation of critical fluctuations associated with electronic charge or orbital nematicity in the tetragonal phase was lacking.

\begin{figure}
\centering
\includegraphics[clip,width=0.99\linewidth]{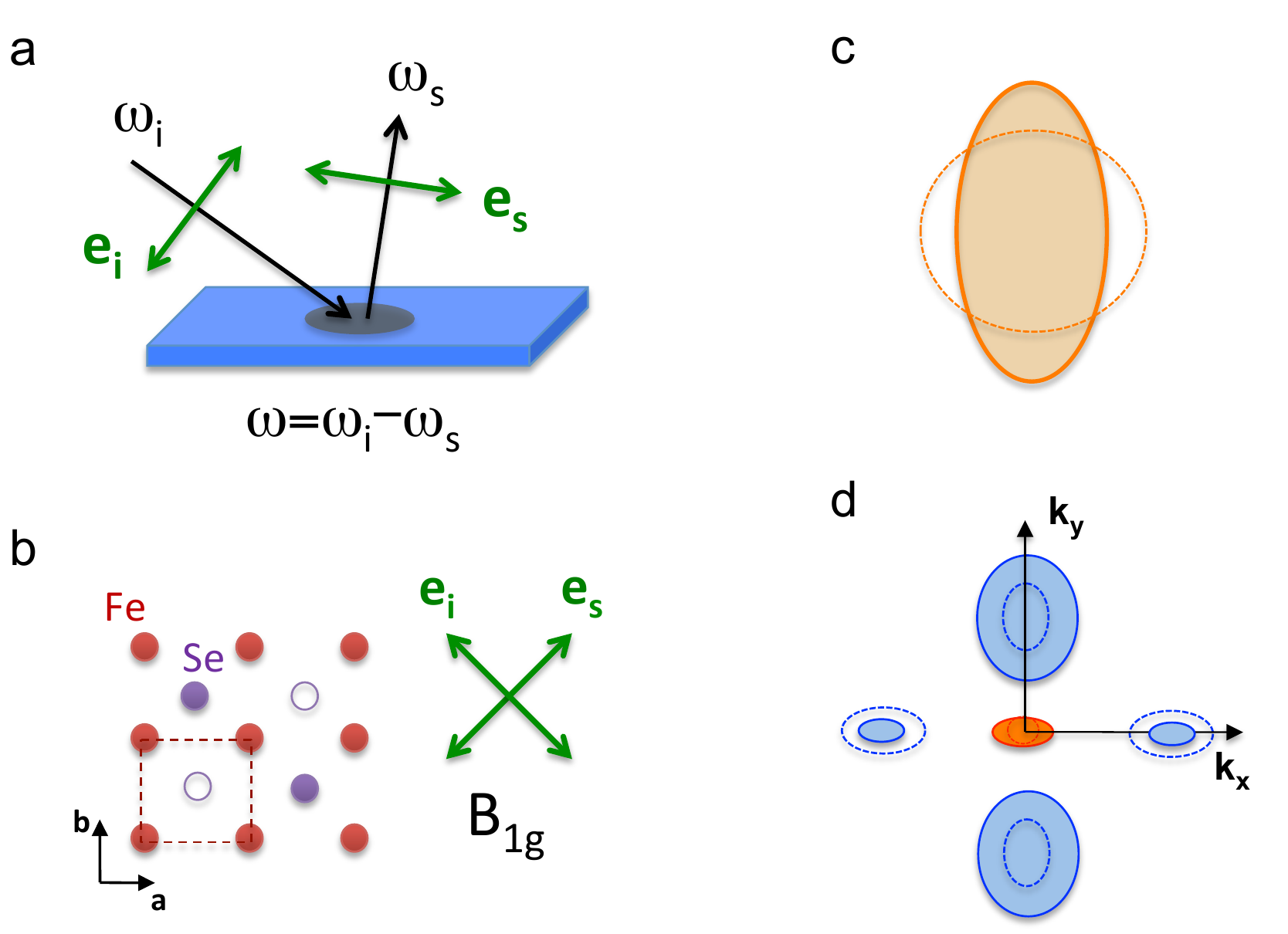}
\caption{(a) Schematic of the Raman scattering process with incoming and scattered photons of frequency $\omega_{i/s}$ and polarization $\textbf{e}_{i/s}$ respectively. The Raman shift is defined as the frequency shift between the incoming and scattered photon frequencies (b) FeSe ab plane with Se atoms alternating above and below the plane defined by the Fe atoms. The 1 Fe unit cell, which neglects the alternating Se atoms, is drawn in dotted lines. In the tetragonal phase above $T_S$, a=b and the crystal structure of FeSe has a fourfold symmetry axis. The $B_{1g}$ symmetry is obtained using crossed incoming and scattered photon polarizations at 45 degrees of the Fe-Fe bonds. (c)-(d) Fermi surface deformation associated to a d-wave Pomeranchuk order for (c) an isotropic Fermi liquid and (d) the multiband Fe SC showing d-wave like deformations with global $B_{1g}$ symmetry which break the fourfold symmetry axis. The deformations shown are consistent with angle resolved photoemission spectroscopy (ARPES) measurements in the orthorhombic phase of FeSe \cite{Suzuki15}: while the hole pocket (red) expand along one direction, the elliptical electron pockets (blue) shrink (expand) along the same (other) direction. The 1 Fe unit cell is used.}
\label{fig1a}
\end{figure}
\par
\begin{figure*}
\centering
\includegraphics[clip,width=0.89\linewidth]{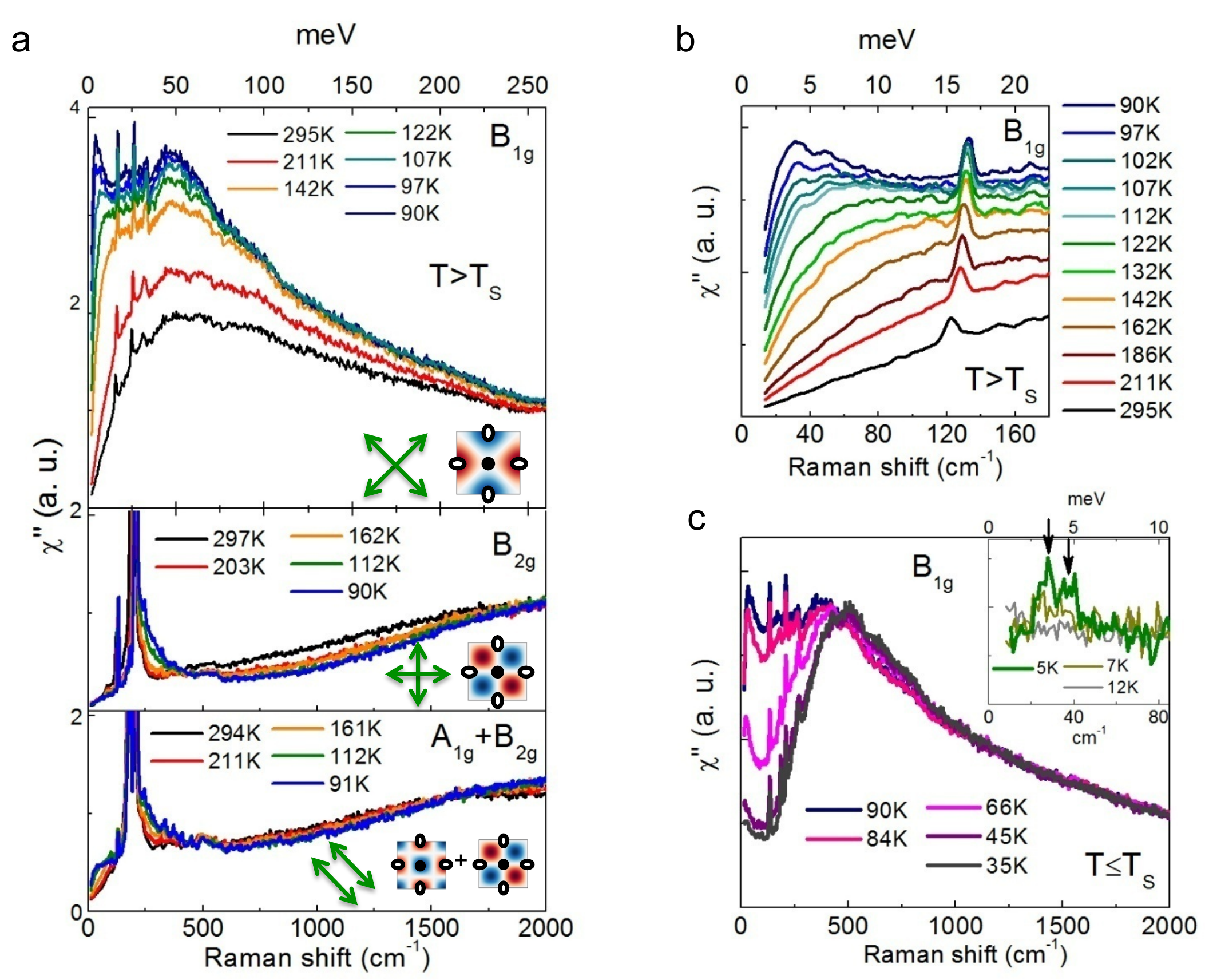}
\caption{(a) Symmetry dependent Raman spectra of FeSe (SP208 crystal) above $T_S$=87~K using 2.33~eV photons. The sharp peaks super-imposed on the electronic continuum are due to Raman active optical phonons. Also shown in inset are the schematics k-space structures of the Raman form factors in different symmetries (blue and red colors indicate positive and negative amplitudes respectively), and the polarization configurations used to select them. (b) Temperature dependence of the low energy $B_{1g}$ spectra above $T_S$.  (c) Evolution of the $B_{1g}$ spectra
accross $T_S$. The inset shows the spectra across the superconducting transition at $T_c$=8.5~K (SP208). The arrows indicate 2$\Delta$ superconducting peaks.}
\label{fig1b}
\end{figure*}
Here, we investigate the nature of nematicity in FeSe by using the unique ability of electronic Raman scattering to selectively probe the dynamics of electronic nematic degrees of freedom free from lattice effects \cite{Yamase13,Gallais13,Kontani14,Gallais-Paul16,Thorsmolle16,Hackl16}. We unravel the presence of critical charge nematic fluctuations in the tetragonal phase which signals the presence of a d-wave Pomeranchuk instability of the Fermi surface \cite{Pomeranchuk58}.
The extracted nematic susceptibility shows quantitative scaling with the measured lattice softening \cite{Fil13,Bohmer15}, demonstrating that charge nematic fluctuations account entirely for the lattice instability. Our results make a strong case for itinerant electronic charge driven nematicity in FeSe.
\par
Raman scattering is a photon-in photon-out process, whereby a monochromatic visible light is inelastically scattered at a different frequency by dynamical fluctuations of the electrical polarizability of the sample (Fig. \ref{fig1a}(a)). In metals the Raman spectra at low frequency shifts are typically composed of sharp optical phonon peaks super-imposed on a broad electronic background, generally referred to as electronic Raman scattering (ERS). The ERS intensity measures the long wavelength dynamical charge correlation function in the symmetry channel $\mu$: $S_{\mu}(\omega) \equiv \langle \rho_{\mu}^{\dagger} (\omega) \rho_{\mu} (\omega) \rangle$, where $\omega$ is the frequency (or Raman) shift between incoming and scattered photons and $\rho_{\mu}$ is the form factor weighted electronic charge \cite{Devereaux2007}. The fluctuation-dissipation theorem in turn links the measured correlation function $S_{\mu}$ to the imaginary part of the Raman response function $\chi_{\mu}''$: $S_{\mu}(\omega)=\frac{1}{\pi}[1+n_B(\omega,T)]\chi_{\mu}''(\omega)$, where $n_B$ is the Bose function.

\par
Being a symmetry resolved probe of the charge fluctuation dynamics with zero momentum transfer, electronic Raman scattering is ideally suited to detect critical in plane charge nematic fluctuations \cite{Yamase13,Gallais13}. The symmetry of the charge fluctuations $\mu$ probed in a Raman experiment is fixed by the directions of the
incoming and scattered photon polarizations. Of interest here is the $B_{1g}$ symmetry (using 1 Fe/cell notation,
see Fig. \ref{fig1a}(b)), obtained for photons polarized along the diagonals
of the Fe-Fe bonds and which transforms as $k_x^2$-$k_y^2$. The $B_{1g}$ charge nematic
fluctuations probed by Raman are equivalent to a Fermi surface deformation with d-wave symmetry. This electronic instability
was predicted by Pomeranchuk to occur in an isotropic Fermi liquid in which the Fermi surface spontaneously deforms along a
specific direction, breaking rotational symmetry \cite{Pomeranchuk58} (see Fig. \ref{fig1a}(c)). In the context of Fe SC the
$B_{1g}$ Raman response probes the fluctuations associated to a multiband version of a d-wave Pomeranchuk order parameter
which breaks the fourfold symmetry axis (Fig. \ref{fig1a}(d)):
$\rho_{B_{1g}}=\sum_{k,\alpha} f_kn_{k,\alpha}$ where $\alpha$ is the orbital index, $f_k$ a d-wave form factor which
transforms as $k_x^2$-$k_y^2$ and $n_k$ the electron density \cite{Gallais-Paul16}.

\par
Raman scattering experiments were performed on two different FeSe crystals (SP208 and MK, see Supplemental Material and \cite{Karlsson15,Suzuki15}). Figure \ref{fig1b}(a) displays the Raman response
$\chi''_{\mu}$ in different symmetries $\mu$ as a function of temperature
in the tetragonal phase ($T>T_S$) for SP208. For comparison besides the response in $B_{1g}$ symmetry, we also show
the response in $B_{2g}$ and $A_{1g}$ symmetries which transform as $k_x k_y$ and $k_x^2$+$k_y^2$ respectively (see form factors in the insets of Fig. \ref{fig1b}(a)). Upon cooling the $ \mu = B_{1g}$ Raman response displays an overall enhancement over a wide energy range extending up to 2000~cm$^{-1}$. At high temperature the response is dominated by a broad peak, centered around 400~cm$^{-1}$ and whose weight increases on cooling. In addition a relatively sharp peak emerges below 100~cm$^{-1}$: it softens and gains considerably in intensity upon approaching $T_S$ (Fig. \ref{fig1b}(b)). By contrast, the response in the two other configurations is only mildly temperature dependent. The $B_{2g}$ response shows a weak suppression above 500 cm$^{-1}$ and a build-up of spectral weight between 200 and 250 cm$^{-1}$, which likely originates from an interband transition between nearly parallel spin-orbit split hole bands at the $\Gamma$ point \cite{Watson15,Zhang15}. Below $T_S$ the $B_{1g}$ response strongly reconstructs (Fig. \ref{fig1b}(c)): the low energy response is suppressed and there is a weak transfer of spectral weight at higher energy, above 500~cm$^{-1}$, in agreement with a previous Raman study \cite{Lemmens13}. Below $T_c$ superconducting gaps open on the different Fermi pockets (inset of Fig. \ref{fig1b}(c)) giving rise to two sharp peaks at 2$\Delta$=28 ($\pm$1)~cm$^{-1}$ ($\sim$3.5~meV) and 37 ($\pm$2)~cm$^{-1}$ ($\sim$4.6~meV), in broad agreement with Scanning Tunneling Microscopy (STM) measurements \cite{Kasahara14}.
\begin{figure}
\centering
\includegraphics[clip,width=0.99\linewidth]{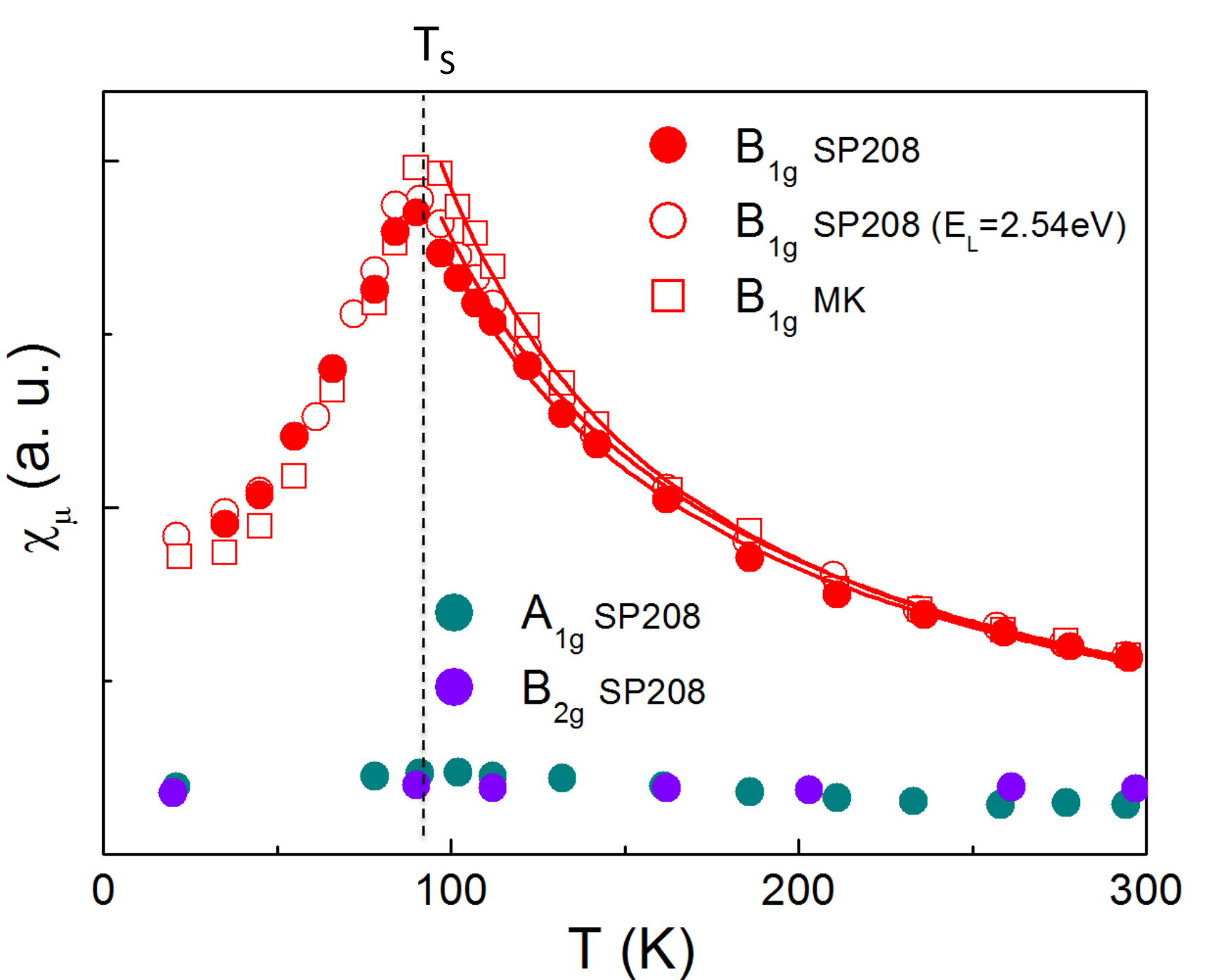}
\caption{Temperature dependence of the $B_{1g}$ charge nematic susceptibility for SP208 ($T_S$=87~K) and MK ($T_S$=88.5~K) using 2.33~eV photons. Also shown are data on SP208 using a different excitation energy (2.54~eV) and the susceptibility in the other symmetry channels on
SP208 ($A_{1g}$ and $B_{2g}$). The lines are Curie-Weiss fits of the $B_{1g}$ susceptibility above $T_S$.}
\label{fig2}
\end{figure}
\par
Focusing on the tetragonal phase, we use the fact that the Raman responses at finite frequency can be
translated into their corresponding symmetry resolved charge susceptibilities at zero-frequency using the Kramers-Kronig relation,
\begin{equation}
\chi_{\mu}(T) =\frac{2}{\pi}\int_0^{\Lambda} \frac{\chi_{\mu}''(T, \omega)}{\omega}d\omega.
\label{KK}
\end{equation}
The susceptibilities obtained
by integrating the finite frequency responses up to $ \Lambda = 2000~cm^{-1}$ are shown as a function of temperature in
Fig. \ref{fig2}. While the $B_{2g}$ and $A_{1g}$ susceptibilities are nearly $T$-independent,
the $B_{1g}$ susceptibility $\chi_{B_{1g}}$ shows a strong enhancement with lowering temperature and
subsequently collapses below $T_S$. This demonstrates the
growth of charge nematic fluctuations in the tetragonal phase which are arrested by the structural transition
at $T_S$. For both SP208 and MK crystals the temperature dependence of $\chi_{B_{1g}}$ above $T_S$ is well captured
by a Curie-Weiss law
$\chi_{B_{1g}}(T) =\frac{B}{T-T_0}$, with a Curie-Weiss temperature $T_0$ significantly below $T_S$,
namely 8~K and 20~K for SP208 and MK respectively.

\par
A key step in the data interpretation is that the nematic fluctuations described above are
entirely electronic in origin, and are not affected by the fluctuations of the orthorhombic strain $u_{xx} - u_{yy}$,
where $\hat{u}$ is the lattice strain tensor \cite{Gallais-Paul16}. The lattice fluctuations are coupled to the electronic
Pomeranchuk order parameter $\rho_{B_{1g}}$ via the electron-phonon interaction
$\mathcal{H}_{\rm el-ph} = \lambda \rho_{B_{1g}} (u_{xx} - u_{yy})$, where $\lambda$ is the coupling constant.
The full, measured nematic susceptibility at momentum ${\bf q}$ along the relevant high-symmetry direction
and frequency $\omega$ can be expressed as
\begin{equation}
\label{full-susceptibility}
(\chi_{B_{1g}})^{-1} ({\bf q}, \omega) = (\chi^0_{B_{1g}})^{-1} ({\bf q}, \omega) - \frac{\lambda^2 q^2}{C_S^0 q^2 - \omega^2}.
\end{equation}
Here $\chi^0_{B_{1g}}({\bf q}, \omega)$ is the electronic susceptibility associated with $\rho_{B_{1g}}$ in the absence of the
lattice, and the second term is the contribution of the orthorhombic strain with the elastic shear modulus $C_S^0$.
Crucially, the nematic susceptibility obtained from the finite frequency Raman spectra ($\omega >$ 8 cm$^{-1}$) using Eq.~(\ref{KK}) is in the dynamical limit,
i.e., $\chi_{\mu}(T) =\lim_{\omega \rightarrow 0} \chi_{\mu} (T, \omega ,  q =0)$. In this limit
the second term of Eq.~(\ref{full-susceptibility}) vanishes, implying that
the extracted nematic susceptibility does not couple to the orthorhombic strain fluctuations~\cite{Kontani14,Gallais16,Gallais-Paul16} and,  therefore, $T_0$ represents the
bare electronic charge nematic transition temperature that is un-renormalized by the lattice.
We conclude that the observed
Curie-Weiss behavior demonstrates the presence of a d-wave Pomeranchuk instability of purely electronic origin in FeSe.
This is in agreement with a recent renormalization group analysis, which shows that the leading instability is in the
Pomeranchuk channel in low Fermi energy systems like FeSe \cite{Chubukov16}.  The d-wave Pomeranchuk order may explain
the peculiar k-dependent orbital splitting observed by angle resolved photoemission spectroscopy (ARPES) below $T_S$,
 which does not fit a simple ferro-orbital order \cite{Watson15,Zhang15,Suzuki15}.

\begin{figure}
\centering
\includegraphics[clip,width=0.99\linewidth]{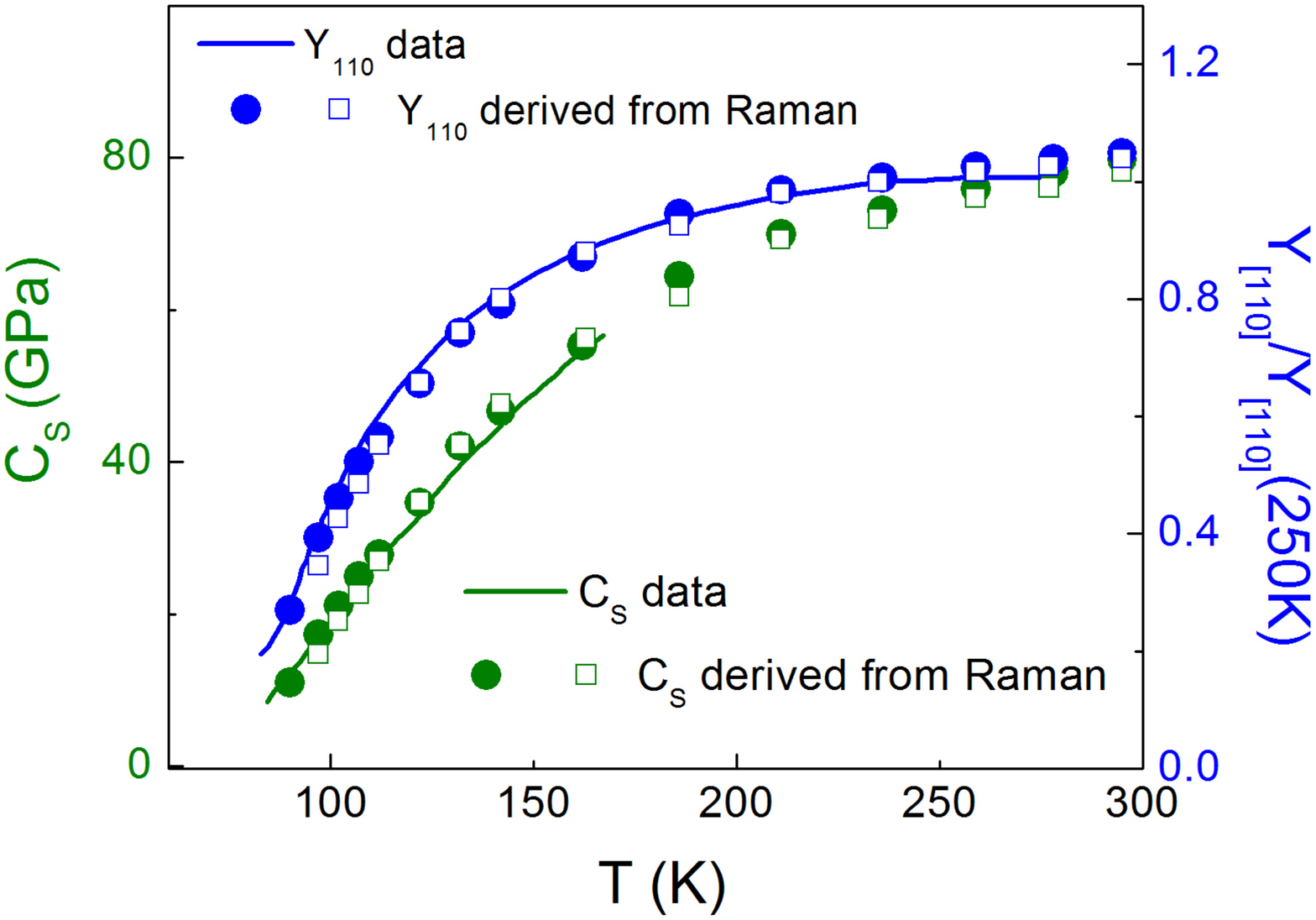}
\caption{Shear modulus $C_{S}$ \cite{Fil13} and Young's modulus $Y_{[110]}$ \cite{Bohmer15} data (line) and corresponding simultaneous fits using the nematic susceptibility $\chi_{B_{1g}}$ extracted from Raman scattering using eq.~\ref{KK} and \ref{Raman-shear}. Full / open symbols correspond to Raman data on SP208 / MK crystal. The $\lambda$ values (in relative units) used for the two crystals agree within 10 percent. The standard relationship between $Y_{[110]}$ and $C_{S}$ was used \cite{Bohmer15} (see Supplements for details).}
\label{fig3}
\end{figure}

\par
Having established the presence of critical charge nematic fluctuations,
we proceed to show that the structural instability at $T_S$ is entirely driven by the reported charge nematic softening.
The renormalization of the relevant shear modulus $C_S$ due to the above-mentioned symmetry-allowed electron-lattice coupling
is given by~\cite{Bohmer15,Gallais-Paul16}
\begin{equation}
C_{S}(T) =C_S^0-\lambda^2 \chi_{B_{1g}}(T),
\label{Raman-shear}
\end{equation}
We take $C^0_S$, the bare modulus, to be $T$-independent as expected for a purely electronic driven structural transition
thus leaving $\lambda$ as the only free parameter. As shown in Fig.~\ref{fig3}, we
find an excellent agreement between the observed softening of $C_S$, obtained either directly from ultra-sound
measurements \cite{Fil13}, or indirectly from Young's modulus measurements \cite{Bohmer15}, and $\chi_{B_{1g}}(T)$
obtained from our Raman measurements. Together with the absence of scaling between elastic modulus and spin
fluctuations, our result makes a strong case for a lattice distortion in FeSe induced by a
d-wave Pomeranchuk instability of the Fermi surface.

\begin{figure*}
\centering
\includegraphics[clip,width=0.99\linewidth]{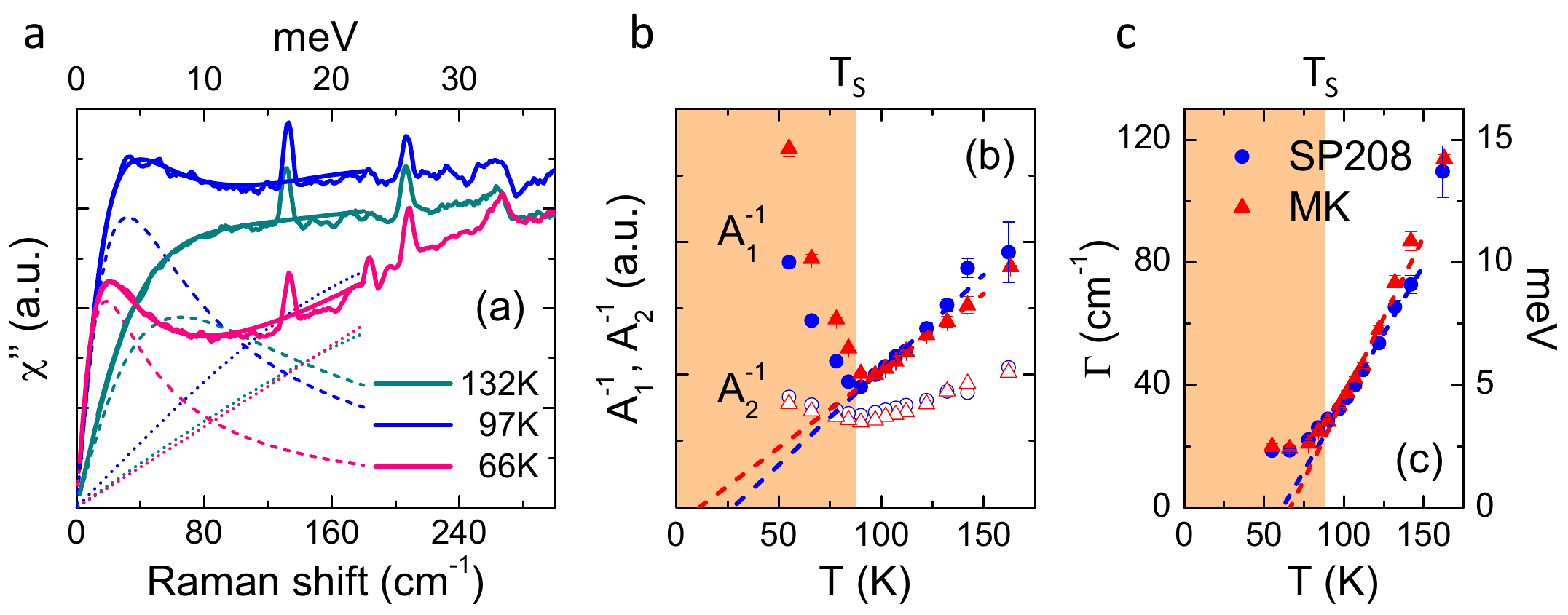}
\caption{(a) Low energy fits of the $B_{1g}$ response of SP208 using a damped Lorentzian for the quasi-elastic peak (QEP)
and an odd in frequency third order polynomial for the low energy part of the broad peak (see Supplements) .
(b) Temperature dependence of the inverse of the two contributions to the nematic susceptibility, $A_{1}^{-1}$ and $A_{2}^{-1}$ for SP208 (blue dots) and MK (red triangle). The dashed line is a linear fit of $A^{-1}_1$ between $T_S$ and 150~K. (c) Temperature dependence of the QEP line width $\Gamma_1$. The dashed line is a linear fit between $T_S$ and 150~K}
\label{fig4}
\end{figure*}

\par
Next, we discuss the frequency dependence of the $B_{1g}$ response in the tetragonal phase. As is evident from the spectra close to $T_S$ in Fig.~\ref{fig1b}(a), the $B_{1g}$ response is composed of two contributions,
a sharp quasi-elastic peak (QEP) at low energy (below 200~cm$^{-1}$),
and a much broader peak centered around 400~cm$^{-1}$. Both features appear only in the $B_{1g}$ symmetry: $\chi''_{B_{1g}}(\omega) = \chi''_{QEP}(\omega) + \chi''_{b}(\omega)$. The low energy QEP is well reproduced by
a damped Lorentzian $\chi''_{QEP}(\omega) =A_1\frac{\omega\Gamma}{\omega^2+\Gamma^2}$, which allows a clear separation of the two contributions, and the extraction of the broad $\chi''_{b}(\omega)$ close to $T_S$ (see Fig. \ref{fig4}(a) and Supplements).
As shown in Fig.~\ref{fig4}(b), their respective contributions $A_1(T)$ and $A_2(T)$ to the nematic
susceptibility $\chi_{B_{1g}}(T)$, through Eq.~(\ref{KK}), have different behavior close to $T_S$ in the tetragonal phase. Only the QEP contribution is critical, with $A_1(T)^{-1}$ extrapolating to zero close to $T_0$. In contrast, the broad peak contribution $A_2(T)$, while sizable, increases only mildly upon cooling. In addition, the extracted QEP line width $\Gamma(T)$ shows a strong softening and extrapolates to zero at $\sim$65~K (Fig.~\ref{fig4}(c)).
\par
In a weak coupling description of a $d$-wave Pomeranchuk instability, the QEP can be understood
as the standard Drude contribution to the Raman conductivity  $\chi''_{B_{1g}}(\omega)/\omega$ with weight $A_1$ and width $\Gamma$ that are
renormalized by the diverging nematic correlation length $\xi$~\cite{Gallais-Paul16}.
Defining $r_0 \equiv \xi^{-2} \propto (T - T_0)$, this theory
predicts $A_1^{-1} \propto r_0$, and $\Gamma \propto \Gamma_0 r_0$, where $\Gamma_0$ is a single particle scattering rate. As shown in Fig.~\ref{fig4} (b) and (c), the linear temperature dependencies of $A_1^{-1}$ and $\Gamma$ between $T_S$ and $T_S+$60~K are in agreement with the above expectation. However, the two quantities extrapolate to zero at different temperatures 20~K ($\pm$ 10~K) and at 65~K ($\pm$ 5~K) respectively. We attribute this mismatch to a strong linear temperature dependence of the scattering rate $\Gamma_0(T)$, as suggested by resistivity measurements \cite{Kasahara14,Karlsson15} (see Supplements).
\par
Finally we discuss the microscopic origin of the broad feature. It is unlikely to be from an Azlamazov-Larkin type
contribution of the fluctuations of the stripe magnetic state \cite{Kontani14,Khodas15,Karahasanovic15,Hackl16} because, below $T_S$, inelastic neutron scattering and NMR data suggest an enhancement of low energy spin fluctuations~\cite{Imai09,Bohmer15,Wang15}, whereas we observe a shift of spectral weight of $\chi''_{b}(\omega)$ to higher frequencies. It is also unlikely that the feature is an interband transition, since
$\chi''_{b}(\omega)$ does not show any gap at low frequencies above $T_S$ (see Supplements). One possibility is that it is the nematic response of electrons that are not sharply defined quasiparticles. Such an interpretation
would be in line with the observed bad metal behavior ~\cite{Kasahara14,Aichhorn10}, and the fact that the Fermi energy of FeSe is rather small ~\cite{Kasahara14,Watson15,Audouard15}.

\par
Overall, our findings support a scenario in which the nematic transition of FeSe is due to
an incipient d-wave Pomeranchuk instability of the Fermi surface. This provides an alternative route to nematicity compared to the prevailing spin fluctuation mediated scenario that has been proposed for other Fe SC. The subsequent challenge will be to identify the microscopic interaction that is responsible for the Pomeranchuk instability, and to study if such an interaction is relevant for other Fe SC as well.
\par
{\bf Materials and Methods} Single crystals of FeSe were grown using chemical vapor transport method based on the use of an eutectic mixture of AlC$_3$/KCl as described in \cite{Karlsson15,Kasahara14}. The two different single crystals measured were grown in Grenoble (SP208) and Kyoto (MK). 
Polarization-resolved Raman experiments have been carried out using a diode-pumped solid
state (DPSS) laser emitting at 2.33~eV. For low energy ($<$ 500 cm$^{-1}$) measurements a triple grating spectrometer
equipped with 1800 grooves/mm gratings and a nitrogen cooled CCD camera were used. Measurements at higher energies, up to 2000~cm$^{-1}$, were performed using a single grating spectrometer with 600 grooves/mm in combination with an ultra-steep edge filter (Semrock) to block the stray-light. Additional measurements were also performed using the 2.54~eV line of an Ar-Kr Laser.

\acknowledgments
We thank G. Blumberg, V. Brouet, A. V. Chubukov, V. D. Fil, R. Hackl and J. Schmalian for fruitful discussions. P. M., M.A. M., M. C., A. S. and Y. G. acknowledge financial support from ANR Grant "Pnictides",  from a Labex SEAM grant and from a SESAME grant from r\'egion Ile-de-France. P. T. and S. Karlsson acknowledge the financial support of  UJF (now integrated inside "Universit\'e Grenoble Alpes") and Grenoble INP through the AGIR-2013 contract of S. Karlsson. S. Kasahara., T. S. and Y. M. acknowledge the support of Grants-in-Aid for Scientific Research (KAKENHI) from Japan Society for the Promotion of Science (JSPS), and  the `Topological Quantum Phenomena' (No. 25103713) Grant-in-Aid for Scientific Research on Innovative Areas from the Ministry of Education, Culture, Sports, Science and Technology (MEXT) of Japan.


\newpage

\textbf{\Large{\center{
Supplementary Information for "Charge induced nematicity in FeSe"
}}}


\section{Methods}

Single crystals of FeSe were grown using chemical vapor transport method based on the use of an eutectic mixture of AlC$_3$/KCl as described in \cite{Karlsson15,Kasahara14}. The two different single crystals measured were grown in Grenoble (SP208) and Kyoto (MK). The structural transition temperatures were determined in-situ by monitoring the emergence of Rayleigh scattering by orthorhombic domains yielding $T_S$=87~K and $T_S$=88.5~K for SP208 and MK respectively (see below for details). These values are in agreement with $T_S$ values extracted from transport measurement (see Fig. \ref{SI5}(a) and \cite{Karlsson15,Kasahara14}.) The superconducting transition temperatures $T_c$ were measured using SQUID magnetometry giving $T_c$=8.5~K (SP208) and $T_c$=9.1~K (MK). Again, these $T_c$ values confirm the values extracted from transport measurements (see Fig. \ref{SI5}(a) and \cite{Karlsson15,Kasahara14}). The crystals were cleaved and transferred to a close-cycle cryostat in inert atmosphere to prevent surface degradation.

\par
Polarization-resolved Raman experiments have been carried out using a diode-pumped solid
state (DPSS) laser emitting at 2.33~eV. For low energy ($<$ 500 cm$^{-1}$) measurements a triple grating spectrometer
equipped with 1800 grooves/mm gratings and a nitrogen cooled CCD camera were used. Measurements at higher energies, up to 2000~cm$^{-1}$, were performed using a single grating spectrometer with 600 grooves/mm in combination with an ultra-steep edge filter (Semrock) to block the stray-light. Additional measurements were also performed using the 2.54~eV line of an Ar-Kr Laser. The laser spot dimension was $\sim$ 50x80~$\mu m^2$. The typical laser power used was 8~mW, but for spectra in the superconducting state laser power less than 0.2~mW was used. All temperatures were corrected for the estimated laser heating (see below for details). 
\par
The $B_{1g}$ and $B_{2g}$ symmetries were obtained using perpendicular incoming and outgoing photon polarizations at 45 degrees, and along of the Fe-Fe bonds respectively. When using parallel incoming and outgoing photon polarizations at 45 degrees of the Fe-Fe bonds, $A_{1g}$ + $B_{2g}$ symmetries are probed. The $A_{1g}$ component can be isolated from the $A_{1g}$+$B_{2g}$ spectra by subtracting the $B_{2g}$ contribution obtained independently.
A piezo-rotator was used to change the orientation of the crystal in-situ with respect to the photon polarizations. In order to extract the symmetry dependent nematic susceptibility from the Raman response at finite frequency using Kramers-Kronig relation, the responses were extrapolated linearly from the lowest frequency measured (8 - 9 cm$^{-1}$ depending on the symmetry and sample) to zero frequency.

\section{Laser heating and determination of $T_S$}

A clear manifestation of the structural transition is the appearance
of Rayleigh scattering at the surface of the crystal due to twin domains formation at $T_{\mathrm{S}}$. This effect is easy to monitor using a camera to visualize the laser spot during Raman experiments. Moreover,
it is very useful to estimate the actual value of $T_{\mathrm{S}}$,
as well as laser heating in-situ. 
\par
\begin{figure*}
\includegraphics[width=1.9\columnwidth]{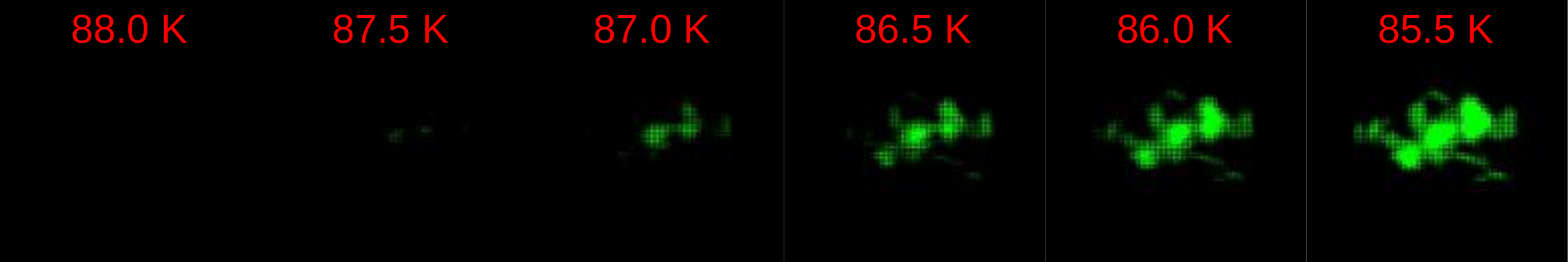}

\protect\caption{Images of the laser spot at different temperatures, taken on an SP208
sample, for a laser power of $\unit[0.5]{mW}$. \label{fig:DS_images}}
\end{figure*}
\begin{figure}[h]
\includegraphics[width=0.9\columnwidth]{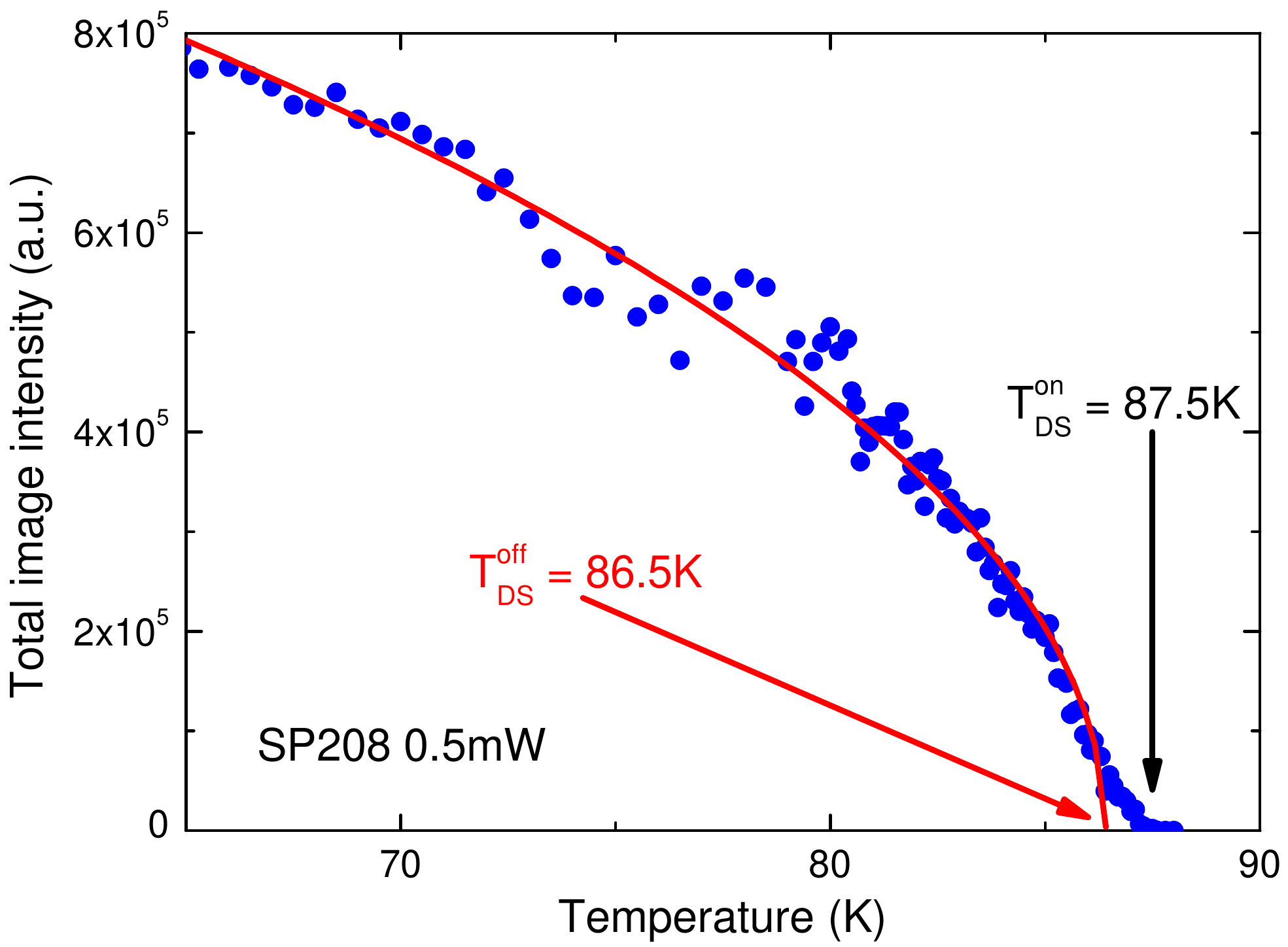}

\protect\caption{Integrated laser spot image intensity as a function of temperature,
for sample SP208, at a laser power of $\unit[0.5]{mW}$. \label{fig:DS_intensity}}
\end{figure}

To achieve this, we take pictures of the laser spot at different temperatures
for a given value of laser power $P_{\mathrm{L}}$ (see figure \ref{fig:DS_images}),
then integrate out the whole spot intensity and plot it as a function
of temperature (see figure \ref{fig:DS_intensity}). 

Figure \ref{fig:DS_images} shows images of the laser spot taken at
different temperatures on the SP208 sample, for a laser power
of $\unit[0.5]{mW}$. These images show the onset of twin domains
scattering at a temperature $T_{\mathrm{DS}}$ between $\unit[87.5]{K}$
and $\unit[87.0]{K}$. Since twin domains scattering appears when
the effective temperature equals $T_{\mathrm{S}}$, the onset temperature
$T_{\mathrm{DS}}$ depends on the value of laser heating. 

Figure \ref{fig:DS_intensity} shows the temperature dependence  of the integrated intensity
 on SP208 at a laser power of $\unit[0.5]{mW}$. It shows an order parameter-like behaviour, which
extrapolates at a temperature $T_{\mathrm{DS,1}}^{\mathrm{off}}=\unit[86.5]{K}$.
Note that the integrated intensity starts to be non-zero at a temperature slightly above
$T_{\mathrm{DS,1}}^{\mathrm{on}}=\unit[87.5]{K}$. This behavior
may be due to either the gaussian tail of the laser spot, for which
the actual laser heating is lower than that measured at the center of
the laser spot, or to a slightly inhomogeneous distribution of $T_{\mathrm{S}}$. 
\par
The same measurement was also performed for a higher
laser power of $\unit[5]{mW}$. The order parameter fit gives $T_{\mathrm{DS,2}}^{\mathrm{off}}=\unit[82.6]{K}$. 
Assuming the following linear relation between the three quantities
$T_{\mathrm{S}}$, $T_{\mathrm{DS}}^{\mathrm{off}}$ and $H_{\mathrm{L}}(T_{\mathrm{S}})$
: 
\[
T_{\mathrm{S}}=T_{\mathrm{DS}}^{\mathrm{off}}+H_{\mathrm{L}}(T_{\mathrm{S}})\times P_{\mathrm{L}}
\]

we can thus determine the actual $T_{\mathrm{S}}$ of our sample and
the laser heating at transition $H_{\mathrm{L}}(T_{\mathrm{S}})$.
We deduce from our two measurements : 
\[
\begin{cases}
T_{\mathrm{S}}=\unit[86.9\pm0.4]{K}\\
H_{\mathrm{L}}(T_{\mathrm{S}})=\unit[0.9]{K/mW}
\end{cases}
\]
The same procedure was applied to sample MK yielding, within error bars, the same estimation of laser heating, but a slightly higher $T_S$: $T_S$=88.5 $\pm$0.5~K.  
Knowing these quantities and the temperature dependence of thermal
conductivity $\kappa(T)$, we can compute an estimation of laser heating
as a function of temperature $H_{\mathrm{L}}(T)$ using the
method described in ref \cite{Maksimov1992}.

\section{Comparison with Co-Ba122}
\begin{figure*}
\includegraphics[width=13cm,trim=0 0 0 0]{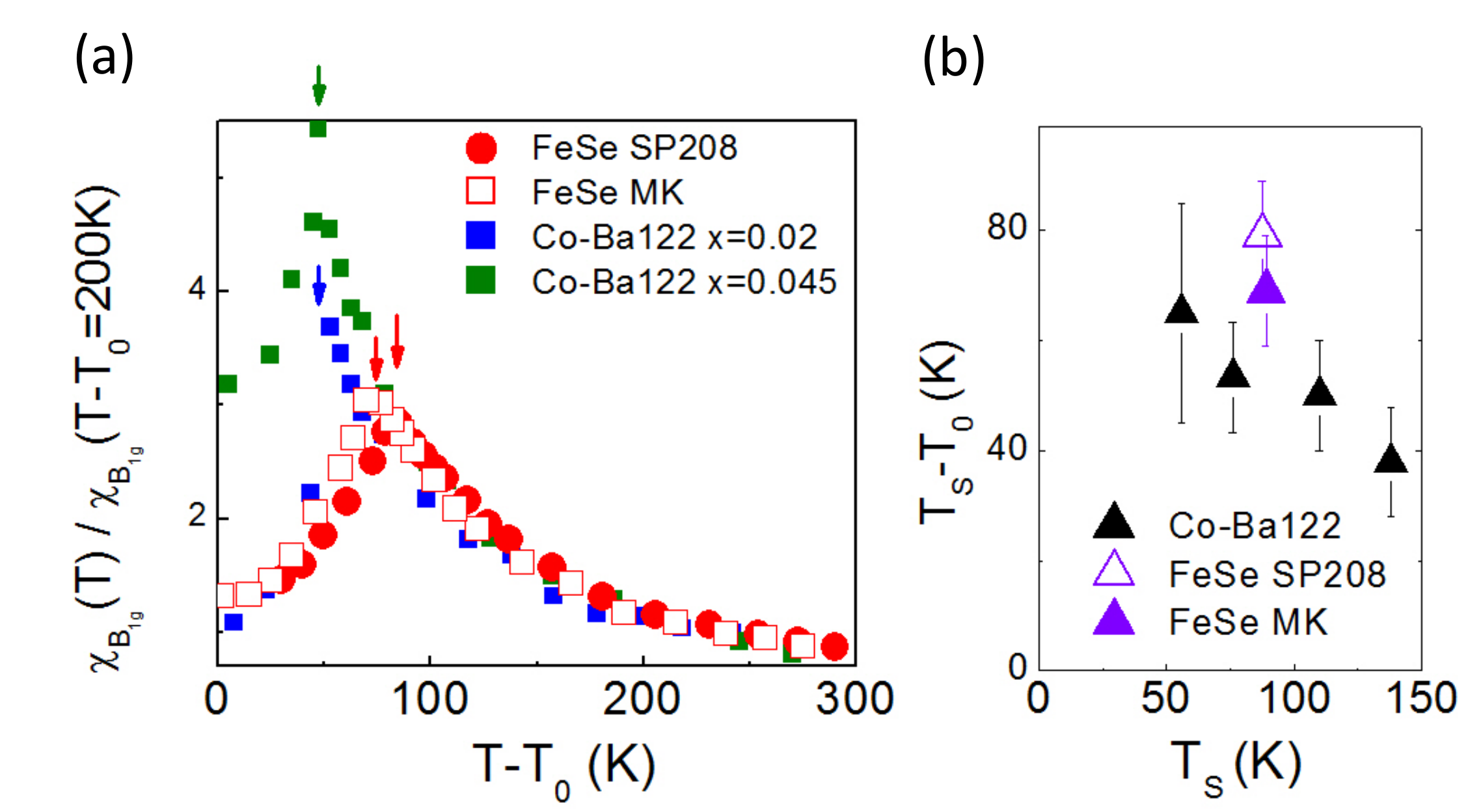}
\caption{(a) Comparison between the $B_{1g}$ charge nematic susceptibility of FeSe and Co doped BaFe$_2$As$_2$ \cite{Gallais13} plotted as a function of T-$T_0$ where $T_0$ is the Curie-Weiss temperature of each sample. The arrows indicate the structural transition $T_S$ for each sample. (b) Electron-lattice coupling energy $T_S$-$T_0$ as a function of $T_S$ for FeSe and Co-Ba122.}
\label{SI4}
\end{figure*}

While there is an apparent contrast between FeSe and other Fe SC with respect to spin degrees of freedom we show in figure \ref{SI4}(a) that, when plotted as a function of $T$-$T_0$, the temperature dependence of the charge nematic susceptibility of FeSe is remarkably similar to the one of electron doped Co-Ba122 \cite{Gallais13}. However the two systems differ in the magnitude of the splitting between $T_S$ and $T_0$, which in a simple Landau-type picture measures the strength of the electron-lattice coupling \cite{Yoshizawa12}. The splitting is 70-80~K for FeSe while it is  less than 60~K in Co-Ba122 (40~K for undoped Ba122) indicating a larger electron-lattice coupling energy in FeSe (Fig. \ref{SI4}(b)).

\section{Link between $C_S$, $Y_{[110]}$ and nematic susceptibility}

Here we give details on the link between the Raman charge nematic susceptibility, the shear modulus and Young's modulus along the [110] direction, as measured in 3-point bending measurements \cite{Bohmer15}. 

\subsection{Link between $C_S$ and $Y_{[110]}$ }

\begin{figure}
\includegraphics[width=8cm,trim=0 0 0 0]{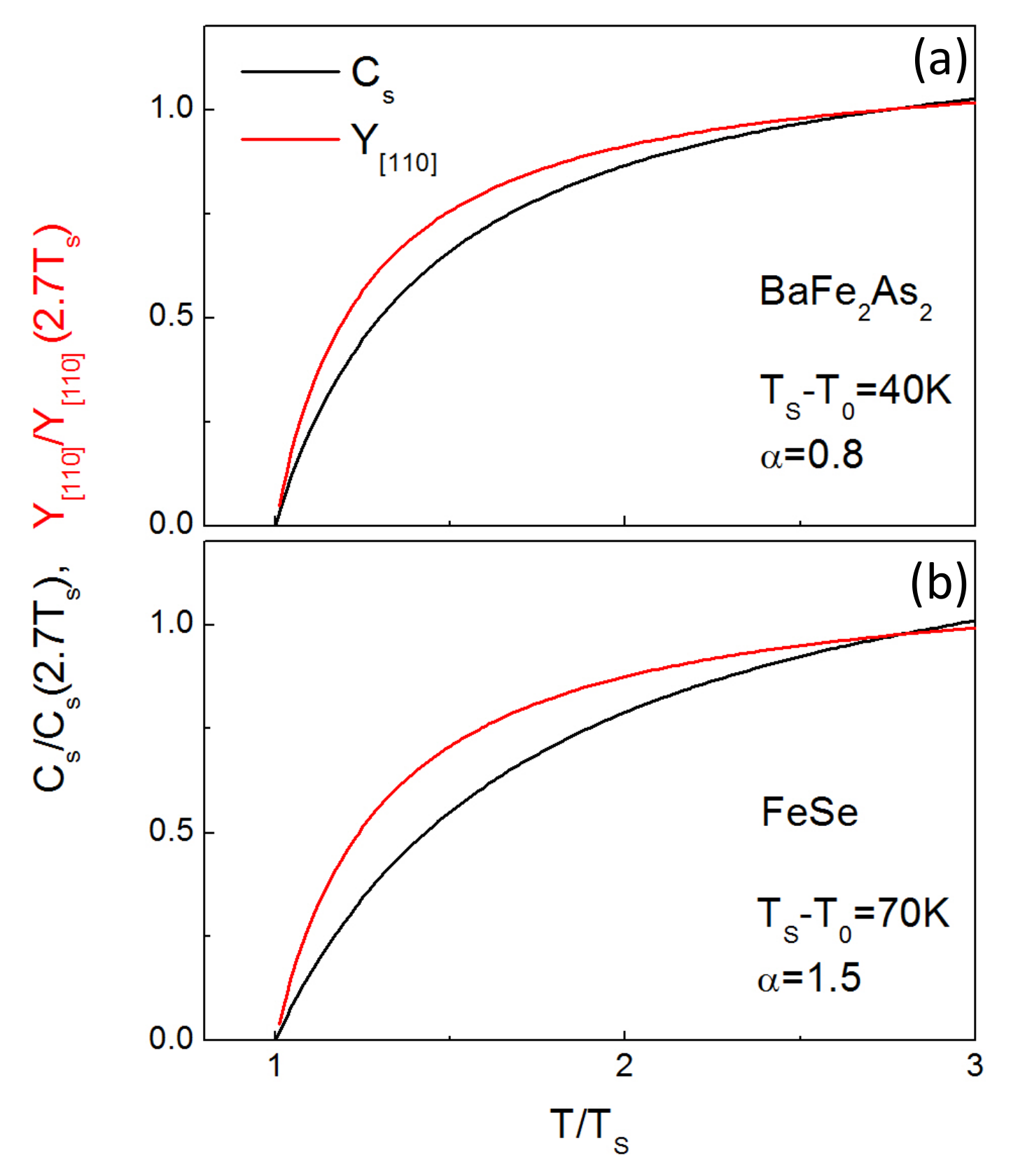}
\caption{Theoretical Curie-Weiss temperature dependences of $C_{S}$ and $Y_{[110]}$ for different value of $T_S-T_0$ and $\alpha$. (a) corresponds to parameters relevant for BaFe$_2$As$_2$ and (b) for FeSe. Values were rescaled at T=2.7T$_S$.}
\label{SI3}
\end{figure}

According to elasticity theory $Y_{[110]}$ can be expressed as a function of the components of the elastic tensor as:
\begin{equation}
Y_{[110]}=4[\frac{1}{C_{S}}+\frac{1}{\gamma}]^{-1}
\label{Y110}
\end{equation}
Using the Voigt notation and the tetragonal (2 Fe) unit cell, the shear modulus is given by $C_S$=$C_{66}$.
The coefficient $\gamma$ depends on four other components: $\gamma=\frac{C_{11}}{2}+\frac{C_{12}}{2}+\frac{C_{13}^2}{C_{33}}$. As in all Fe SC, $C_{S}$ is the only soft component above $T_S$. The other components have a weak temperature dependence due to anharmonicity and can be safely approximated as constants between 300K and $T_S$ \cite{Yoshizawa14,Fil13}. Sufficiently close to $T_S$ the behavior of $Y_{[110]}$ will therefore be dominated by $C_{S}$ which strongly softens: $\frac{Y_{[110]}}{Y^0_{[110]}}\approx \frac{C_{S}}{C_S^0}$. However this approximation holds only very close $T_S$, and far away from $T_S$ the proportionality between the 2 quantities will not be verified. This can be illustrated by assuming that $C_{S}$ follows a Curie-Weiss dependence as observed in Co-Ba122:
\begin{equation}
C_{S}=C_{S}^0\frac{T-T_S}{T-T_0}
\end{equation}
$T_S$ is the structural transition temperature and $T_0$ can be thought as the electronic nematic transition temperature in the absence of coupling to the lattice. $T_0$ can be identified as the Curie-Weiss temperature extracted from Raman measurements if only charge and lattice degrees of freedom are considered.
It is straightforward to compute the corresponding temperature dependence of $Y_{[110]}$:
\begin{equation}
Y_{[110]}=\frac{4C_{S}^0}{1+\alpha}\frac{T-T_S}{T-T_1}
\end{equation}
where $\alpha=\frac{C_{S}^0}{\gamma}$ and 
\begin{equation}
T_1=\frac{T_0+\alpha T_S}{1+\alpha}
\end{equation}
The temperature dependence of $Y_{[110]}$  is still of Curie-Weiss type but with a new characteristic temperature $T_1 \neq T_0$. In the limit where $\alpha<<1$ we have $T_1\approx T_0$. This limit is however never reached in Fe SC where $\alpha\approx$0.7 in Ba122 and $\alpha\approx$=1.5 in FeSe. In general $T_1$ will be bounded by $T_0$ from below, and $T_S$ from above: $T_0<T_1<T_S$. The disagreement between $C_{S}$ and its estimate from $Y_{[110]}$ will therefore be marginal when $T_0$ and $T_S$ are close like in Ba122 (about 40~K). However in FeSe where $T_S-T_0\sim 70~K$ the difference is more significant, and the full expression \ref{Y110} must be used. This is illustrated in fig. \ref{SI3} (a) and (b) where the normalized temperature dependences of $C_{S}$ and $Y_{[110]}$ are plotted for parameters relevant to Ba122 and FeSe respectively. 

\subsection{Simultaneous scaling of $C_{S}$ and $Y_{[110]}$ with $\chi_{B_{1g}}$}
The comparison between the experimentally observed softening of $C_{s}$ and the one expected from the charge nematic susceptibility $\chi_{B_{1g}}$ was performed using equation 2 of the main text with $C_{s}^0$=105 (110)~GPa for SP208 (MK) sample. The only free parameter was the electron-lattice coupling $\lambda$ and good agreement was found for each sample in the temperature interval where $C_{S}$ was measured. The $\lambda$ value used for SP208 was 10\% higher than for MK. Note that since $\chi_{B_{1g}}$ extracted from Raman measurements is only known in relative units, we cannot access the absolute value of $\lambda$ from the fits. Using the same parameters, $C_S^0$ and $\lambda$, the associated softening of $Y_{[110]}$ was then computed using equation \ref{Y110} with $\gamma$=70~GPa, as estimated from both elastic constant measurements and ab-initio calculations \cite{Fil13,Chandra10}. As $Y_{[110]}$ is only known up to a constant prefactor, the data were rescaled at 250~K \cite{Bohmer15}.  
\begin{figure*}
\includegraphics[width=0.8\linewidth]{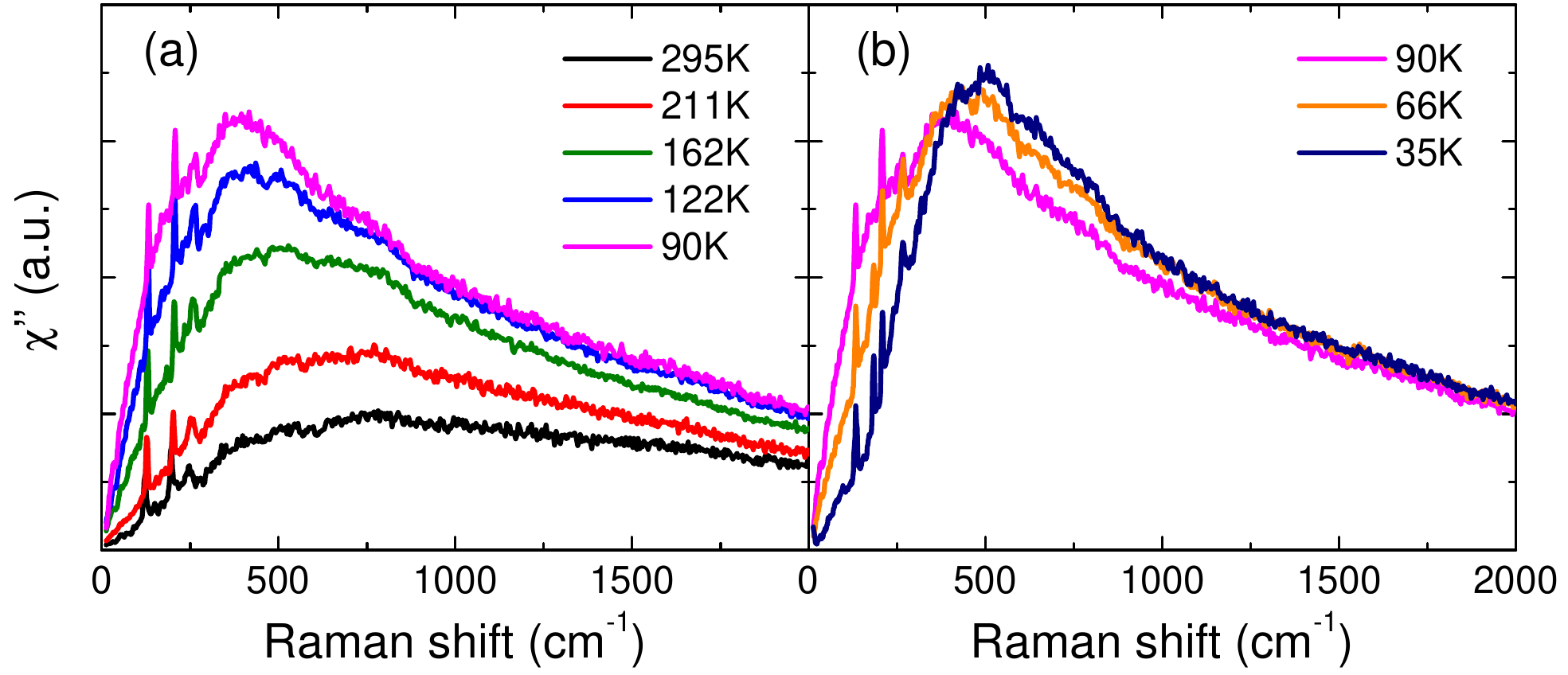} \protect\caption{High energy contribution to the $B_{1g}$ Raman response, with the low energy QEP component subtracted, of sample SP208 at 2.33~eV
($\unit[532]{nm}$) as a function of temperature (a) above $T_{\mathrm{S}}$
and (b) below $T_{\mathrm{S}}$. 
\label{fig:High-energy-contribution}}
\end{figure*}
\section{Fitting procedure of the finite frequency $B_{1g}$ response}

\subsection{Fits using a quasi-elastic peak and a  low energy background}

In order to quantify the temperature dependences of the two components contributing 
to nematic fluctuations in FeSe, it is necessary to fit Raman response
data, especially the low-energy Quasi-Elastic Peak (QEP). To achieve
this, we used the following general expression : 
\begin{equation}
\chi^{''}(\omega,T)=\chi{}_{\mathrm{QEP}}^{''}(\omega,T)+\chi''_b(\omega,T)\label{eq:QEP_odd}
\end{equation}
where the QEP is modeled by an damped Lorentzian: 
\begin{equation}
\chi{}_{\mathrm{QEP}}^{''}(\omega,T)=A_1(T)\frac{\Gamma(T)\omega}{\omega^{2}+\Gamma{}^{2}(T)}
\label{eq:qep}
\end{equation}
At low energy the broad peak $\chi''_b$ was modeled using a third order polynomial form with only odd powers in $\omega$ to guaranty causality. 

\begin{equation}
\chi''_b(\omega,T)= b_{1}(T)\omega+b_{3}(T)\omega^{3}
\end{equation}

\begin{figure*}[h]
\includegraphics[width=0.9\textwidth]{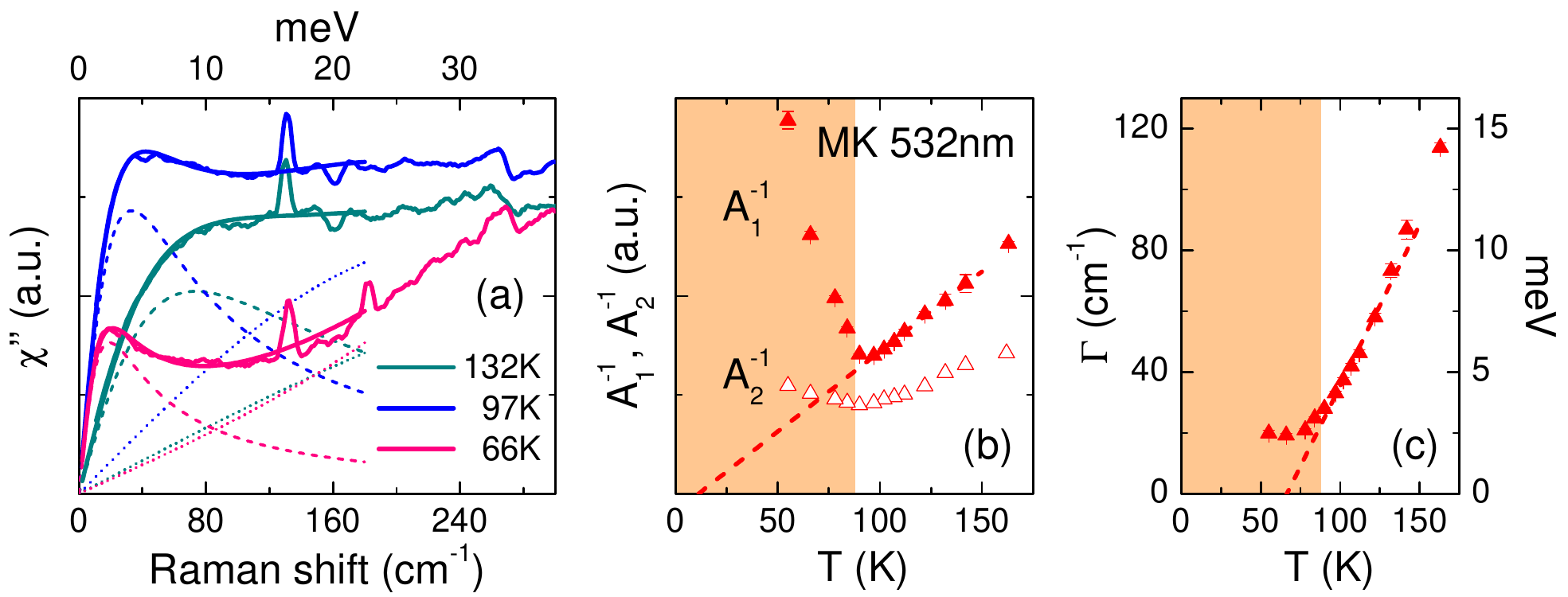}\protect\caption{(a) Low energy fits of the $B_{1g}$ response of sample MK at 2.33~eV ($\unit[532]{nm}$)
using equation (\ref{eq:QEP_odd}). (b) Temperature dependences of
the inverse of the two contributions to the nematic susceptibility,
$A_{1}$ and $A_{2}$ (red triangles, same as in figure 4 of the main
text). The dashed line is a linear fit of $A_{1}^{-1}$ between $T_{\mathrm{S}}$
and $\unit[150]{K}$. It crosses the $x$-axis at $T^{*}=\unit[11]{K}$.
(c) Temperature dependence of the line width $\Gamma$ of the QEP.
The dashed line is a linear fit between $T_{\mathrm{S}}$ and $\unit[150]{K}$.
It crosses the $x$-axis at $T^{**}=\unit[66]{K}$. \label{fig:odd_MK}}
\end{figure*}

\begin{figure*}[h]
\includegraphics[width=0.9\textwidth]{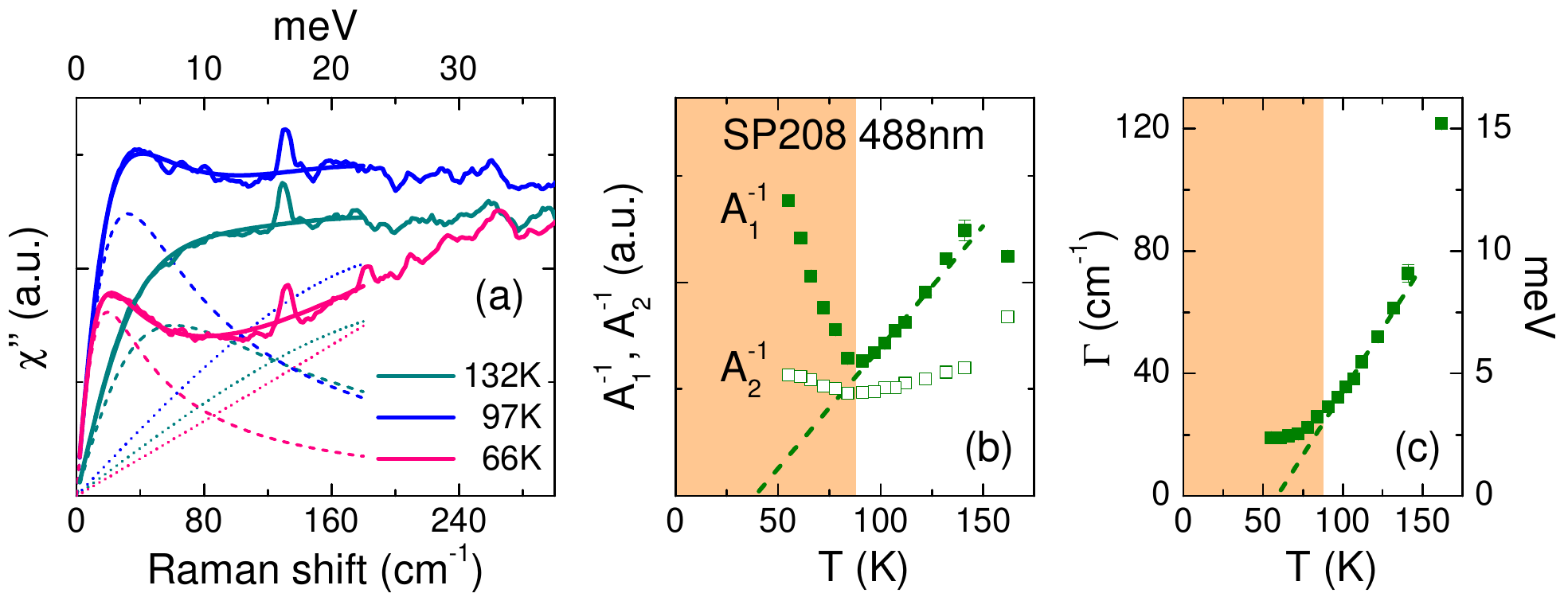}

\protect\caption{(a) Low energy fits of the $B_{1g}$ response of sample SP208 at 2.54~eV ($\unit[488]{nm}$)
using equation (\ref{eq:QEP_odd}). (b) Temperature dependences of
the inverse of the two contributions to the nematic susceptibility,
$A_{1}$ and $A_{2}$. The dashed line is a linear fit of $A_{1}^{-1}$
between $T_{\mathrm{S}}$ and $\unit[150]{K}$. It crosses the $x$-axis
at $T^{*}=\unit[39]{K}$. (c) Temperature dependence of the line width
$\Gamma$ of the QEP. The dashed line is a linear fit between $T_{\mathrm{S}}$
and $\unit[150]{K}$. It crosses the $x$-axis at $T^{**}=\unit[59]{K}$.
\label{fig:odd_488}}
\end{figure*}

As is clear from figure 4(a) of the
main text and from figures \ref{fig:odd_MK}(a) and \ref{fig:odd_488}(a),
equation (\ref{eq:QEP_odd}) fits well the Raman response data at
low energy, up to at least $\unit[180]{cm^{-1}}$, and at all temperatures,
above and below $T_{\mathrm{S}}$. In particular,
below $T_{\mathrm{S}}$, the high energy peak is partially
gapped (see also figure \ref{fig:High-energy-contribution}), resulting in a change in parameter $b_{3}(T)$ from negative
values above $T_{\mathrm{S}}$ to positive values deep below $T_{\mathrm{S}}$.
\par
Figures 4(b) of the main text and \ref{fig:odd_MK}(b) and \ref{fig:odd_488}(b)
show the temperature dependences of the inverse of the two contributions
to the nematic susceptibility, $A_{1}$ and $A_{2}$. The temperature dependences $A_{1}(T)$ for all samples
were fitted between $\unit[95]{K}$ and $\unit[150]{K}$ using a linear
form $A_{1}(T)=a_{1}.(T-T^{*})$.  $A_{2}(T)$ was computed using the following method : the low energy QEP fits
were subtracted from the full Raman responses (Fig. \ref{fig:High-energy-contribution}).
The spectra were then divided by frequency
and integrated up to $\unit[2000]{cm^{-1}}$. 
\par
Figures 4(c) of the main text and \ref{fig:odd_MK}(c) and \ref{fig:odd_488}(c)
show the temperature dependences of the line width $\Gamma$ of the
QEP, directly extracted from fits of the Raman response using equation
(\ref{eq:QEP_odd}). The temperature dependences $\Gamma(T)$
were fitted between $\unit[95]{K}$ and $\unit[150]{K}$
using a linear form $\Gamma(T)=\Gamma_0(T-T^{**})$.

\subsection{Fits using two quasi-elastic peaks}

\begin{figure*}[h]
\includegraphics[width=0.9\textwidth]{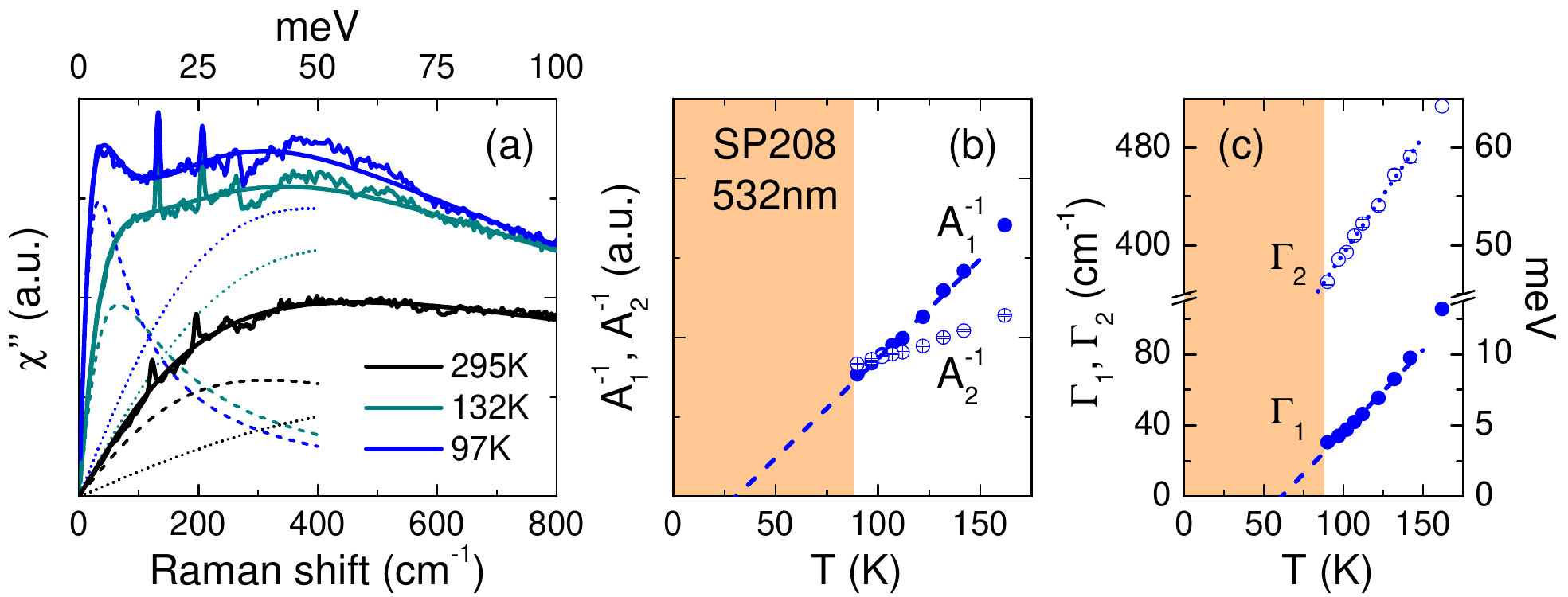}

\protect\caption{(a) Low energy fits of the $B_{1g}$ response of sample SP208 at 2.33~eV ($\unit[532]{nm}$)
using equation (\ref{eq:2QEP}). Note that the energy range is wider
than that of figures 4(a), \ref{fig:odd_MK}(a) and \ref{fig:odd_488}(a).
(b) Temperature dependences of the inverse of the two contributions
to the nematic susceptibility, $A_{1}$ and $A_{2}$. The dashed line
is a linear fit of $A_{1}^{-1}$ between $T_{\mathrm{S}}$ and $\unit[150]{K}$.
It crosses the $x$-axis at $T^{*}=\unit[31]{K}$. (c) Temperature
dependences of the line widths $\Gamma_{1}$ and $\Gamma_{2}$ of
the low and high energy QEPs, respectively. The dashed and dotted
lines are linear fits between $T_{\mathrm{S}}$ and $\unit[150]{K}$.
The dashed line crosses the $x$-axis at $T^{**}=\unit[61]{K}$. \label{fig:2QEP}}
\end{figure*}

As shown in figure \ref{fig:2QEP}(a), we found
that the data can also be well fitted above $T_{\mathrm{S}}$ with
a sum of two QEPs, as expected from the contributions of two intraband, Drude-like, terms: 
\begin{equation}
\chi^{''}(\omega,T>T_{\mathrm{S}})=\chi{}_{\mathrm{QEP}_{1}}^{''}(\omega,T)+\chi{}_{\mathrm{QEP}_{2}}^{''}(\omega,T)\label{eq:2QEP}
\end{equation}
where

\begin{equation}
\chi{}_{\mathrm{QEP}_{1,2}}^{''}(\omega,T)=A_{1,2}(T).\frac{\Gamma_{1,2}(T)\omega}{\omega^{2}+\Gamma_{1,2}^{2}(T)}
\label{eq:qep2}
\end{equation}

Fits using equation (\ref{eq:2QEP}) are good up to $\unit[1000]{cm^{-1}}$, strengthening our interpretation that the brand peak arises from more incoherent intraband excitations. Note however, that because of the partial gapping mentioned
above, the two QEP analysis does not reproduce the data satisfactorily below $T_S$. Figure \ref{fig:2QEP}(b) shows the temperature dependences of the
inverse of the two contributions to the nematic susceptibility, $A_{1}$
and $A_{2}$ extracted from the fits. Figure \ref{fig:2QEP}(c) shows the temperature dependences of the
line widths $\Gamma_{1}$ and $\Gamma_{2}$ of QEP$_{1}$ and QEP$_{2}$,
respectively. Both quantities show linear temperature dependences,
indicated by dashed and dotted lines, respectively. 

\section{QEP line width and resistivity}

In a Random Phase Approximation picture of a d-wave Pomeranchuk transition we expect the QEP amplitude $A_1^{-1}(T)$ to scale as $r_0(T)$. Here $r_0(T)=\xi^{-2} \propto T-T_0$, where $\xi(T)$ is the nematic correlation length and $T_0$  the mean-field nematic transition temperature \cite{Gallais-Paul16}. Experimentally $T^*$, the zero temperature intercept of $A_{1}^{-1}(T)$ ($T^*\sim$25~K ($\pm$ 15~K)), is indeed close to $T_0$, as obtained from the global Curie-Weiss fit of $\chi_{B_{1g}}$.  However the zero temperature intercept of the QEP line width $\Gamma$,  $T^{**}$, is significantly higher: $T^{**}\sim$ 65~K ($\pm$5~K).
\par
Here we show that the shift between $T^*$ and $T^{**}$ can be accounted by the temperature dependence of the bare quasiparticle scattering $\Gamma_0(T)$ as measured by e.g. transport. In FeSe, and in contrast to e.g. BaFe$_2$As$_2$, the resistivity is strongly temperature dependent above $T_S$. Between $T_S$ and 200~K it show quasi-linear behavior with a positive intercept on the temperature axis. Since $\Gamma(T) \propto \Gamma_0$(T) $r_0(T)$, the temperature dependence of the QEP line width $\Gamma$ will contain contributions coming from both the quasiparticle scattering rate $\Gamma_0(T)$ and $r_0(T)$.
\par
\begin{figure*}
\includegraphics[width=16cm,trim=0 0 0 0]{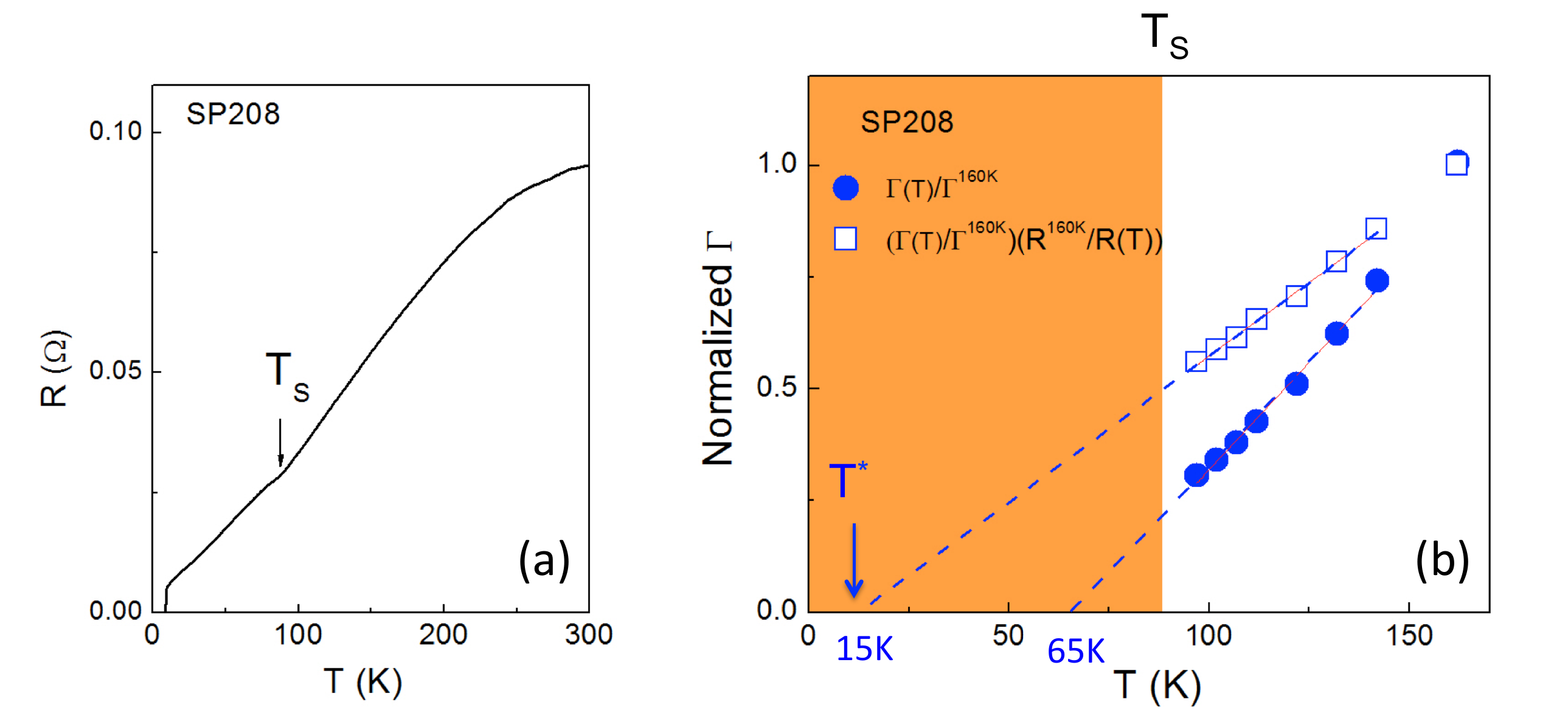}
\caption{(a) Temperature dependence of the resistance $R$ of FeSe SP208. (b) Temperature dependence of $\Gamma$(T) and $\frac{\Gamma(T)}{R(T)}$, normalized at their 160~K values on FeSe SP208}
\label{SI5}
\end{figure*}

Assuming that $\Gamma_0(T)$ is proportional to the resistivity, $\Gamma_0\propto R$, we can extract the temperature dependence of $r_0$(T) by dividing the measurement of QEP line width $\Gamma$ by the resistance $R$:

\begin{equation}
\frac{\Gamma(T)}{R(T)} \propto r_0(T)
\end{equation}
We have used the resistivity data on a crystal from the same batch as SP208 (fig. \ref{SI5}) to correct the temperature dependence of $\Gamma$(T). $\Gamma$(T) and $\frac{\Gamma(T)}{R(T)}$, normalized at their 160~K values, are shown in Fig. \ref{SI5}(b). While the temperature dependence of $\Gamma$(T) between $T_S$ and 150~K extrapolates linearly at $T^{**}\sim$65~K, the quantity $\frac{\Gamma(T)}{R(T)}$ extrapolates at a lower temperature $\sim$ 15~K, now much closer to the value of $T^*$ extracted from the temperature dependence of QEP amplitude $A_1^{-1}$. Taking into account the temperature dependence of the scattering rate $\Gamma_0(T)$ thus reconciles the temperature dependences of $A_1$(T) and $\Gamma$(T).

\end{document}